\documentclass[fleqn,10pt]{wlscirep}
\usepackage{graphicx}
\usepackage{epstopdf}
\usepackage{amsmath}
\usepackage{color}

\begin{document}
\title{Quasiparticle Properties under Interactions in Weyl and Nodal Line Semimetals}
\author[1,*]{Jing Kang}
\author[1]{Jianfei Zou}
\author[2,*]{Kai Li}
\author[3,4]{Shun-Li Yu}
\author[3,4]{Lu-Bing Shao}
\affil[1]{College of Science, Hohai University, Nanjing 210098, China}
\affil[2]{School of Physics and Engineering, Zhengzhou University, Zhengzhou 450001, China}
\affil[3]{National Laboratory of Solid State Microstructures and Department of Physics, Nanjing University, Nanjing 210093, China}
\affil[4]{Collaborative Innovation Center of Advanced Microstructures, Nanjing University, Nanjing 210093, China}
\affil[*]{jkang@hhu.edu.cn,kaili@zzu.edu.cn}
\date{\today}

\begin{abstract}
The quasiparticle spectra of interacting Weyl and nodal-line semimetals on a cubic lattice are studied using the cluster perturbation theory. By tracking the spectral functions under interaction, we find that the Weyl points will move to and meet at a specific point in one Weyl semimetal model, while in the other Weyl semimetal model they are immobile. In the nodal-line semimetals, we find that the nodal line shrinks to a point and then disappears under interaction in one-nodal-line system. When we add another nodal line to this system, we find that the two nodal lines both shrink to specific points, but the disappearing processes of the two nodal lines are not synchronized. We argue that the nontrivial evolution of Weyl points and nodal lines under interaction is due to the presence of symmetry breaking order, e.g., a ferromagnetic moment, in the framework of mean field theory, whereas the stability of Weyl points under interaction is protected by symmetry. Among all these models, the spectral gap is finally opened when the interaction is strong enough.

\end{abstract}
\maketitle

\section*{Introduction}
In recent years, Weyl semimetals\cite{Wan, Balents, Yan, Armitage}(WSM) become more and more attractive since the discovery of the abnormal Fermi arc surface state\cite{Togo}. WSM materials such as the TaAs-family pnictides\cite{Weng,Huang1} and MoTe$_{2}$\cite{Soluyanov,Sun} have been discovered by observation of the fantastic Fermi arcs of surface states through angle-resolved photoemission spectroscopy\cite{Lv, Xu, Yang, Deng, Jiang, Huang2}. The WSMs have a Fermi surface consisting of a finite number of WPs in the Brillioun zone(BZ), at which the conduction and valence bands meet linearly. Such a phase has massless Weyl quasiparticles which can be viewed as half-Dirac Fermions. Besides, the Weyl quasiparticles can be gapped by coupling two quasiparticles with different chirality\cite{William,Turner}. In the presence of interactions which can easily arise in realistic systems, the Weyl quasiparticles can be moved, normalized and even gapped. Another interesting system is the nodal-line semimetal(NLSM) with one-dimensional Fermi surfaces\cite{Burkov,Phillips,Liu,Behrends}. In the experiment side, the NLSMs were observed in several compounds such as PbTaSe$_{2}$\cite{Bian} and ZrSiS\cite{Schoop,Neupane}. In this system there are bulk band touchings along 1D lines and these line-like touchings need extra symmetries to be topologically protected. Interactions can also be applied in this system to discuss the proximity effect and spontaneous symmetry breaking\cite{Roy}.

In this paper, by using the Cluster Perturbation Theory (CPT)\cite{Senechal,Senechal1,Senechal2,Kang,Yu,Li}, we study two lattice models for WSMs and one for NLSM to see how the on-site Coulomb interaction affects the WPs and nodal lines. In CPT, the quasiparticle spectral function can be calculated and then the positions of the WPs and nodal lines can be tracked through the spectral function when the interaction alters. We find that the WPs will move to and meet each other under interactions in one WSM model, while in the other WSM model they will not. In the one-nodal-line semimetal system, the nodal line shrinks to a point under interactions. When we add another nodal line to this system, we find that the two nodal lines will all shrink to specific points one after the other. We argue that the nontrivial evolution of WPs and nodal lines under interaction is due to the presence of symmetry breaking order, e.g., a ferromagnetic moment, in the framework of mean field theory, whereas the stability of WPs under interaction is protected by symmetry. Among all these models, the spectral gap is finally opened when the interaction is strong enough.

\section*{Models and Methods}
\textit{Weyl Semimetal Models:} We will consider two kinds of Weyl Semimetal Models. For the first kind of WSM model (WSM1), the tight-binding Hamiltonian is written as follows\cite{William,Yang1}:
\begin{equation}
H_{0}=\sum_{k}c^{\dag}_{k}\{[2t(\cos k_{x}-\cos k_{0})+m(2-\cos k_{y}-\cos k_{z})]\sigma_{x}+2t\sin k
_{y}\sigma_{y}+2t\sin k_{z}\sigma_{z}\}c_{k},
\label{model1}
\end{equation}
where $c^{\dag}_{k}=(c^{\dag}_{k\uparrow},c^{\dag}_{k\downarrow})$ and $\sigma_{x,y,z}$ are the Pauli matrices. The hopping constant $t$ will be set $t=1$ in calculations based on this model. This model breaks both time-reversal and space-inversion symmetries. After diagonalizing this non-interacting Hamiltonian, we can get the band structure
\begin{equation}
\epsilon_{k}=\pm\sqrt{[2(\cos k_{x}-\cos k_{0})+m(2-\cos k_{y}-\cos k_{z})]^{2}+4\sin^{2}k_{y}+4\sin^{2}k_{z}}.
\end{equation}
Assuming $\epsilon_{k}=0$, we get the WPs in the bulk BZ. There can be 2, 6 or 8 WPs for different parameter $m$, while two of the WPs are present at $(\pm k_{0},0,0)$.

We now introduce the second kind of WSM model (WSM2) which, in contrast to both the WSM1 and the second Weyl semimetal model studied in Ref. \cite{William}, preserves the space-inversion symmetry but not the time-reversal symmetry. The tight-binding Hamiltonian is written as follows:
\begin{equation}
H_{0}=2t\sum_{k}c^{\dag}_{k}[\cos(k_{x})\sigma_{x}+\cos(k_{y})\sigma_{y}+\cos(k_{z})\sigma_{z}]c_{k}.
\label{model2}
\end{equation}
Similar to the second Weyl semimetal model studied in Ref. \cite{William}, the real space hopping is a type of bond-selective and spin-dependent Kitaev-like hopping \cite{Li}. After diagonalizing this non-interacting Hamiltonian, we can get the band structure
\begin{equation}
\epsilon_{k}=\pm 2t\sqrt{\cos^{2}k_{x}+\cos^{2}k_{y}+\cos^{2}k_{z}}.
\end{equation}
This model gives us eight WPs at $(\pm\frac{\pi}{2},\pm\frac{\pi}{2},\pm\frac{\pi}{2})$.

\textit{Nodal-Line Semimetal Model:} For the NLSM model, the tight-binding Hamiltonian is written as follows\cite{Roy}:
\begin{equation}
H_{0}=\sum_{k}c^{\dag}_{k}\{2[t_{1}(\cos k_{x}+\cos k_{y}-b)+t_{2}(\cos k_{z}-1)]\sigma_{x}+2t_{3}\sin k_{z}\sigma_{y}\}c_{k},
\label{model3}
\end{equation}
where $t_{1}$,$t_{2}$,$t_{3}$ are the hopping constants. We can also diagonalize the non-interacting Hamiltonian to get the band structure, which reads:
\begin{equation}
\epsilon_{k}=\pm 2\sqrt{[t_{1}(\cos k_{x}+\cos k_{y}-b)+t_{2}(\cos k_{z}-1)]^{2}+t_{3}^{2}\sin^{2}k_{z}}.
\end{equation}
During our calculations based on this model, the hopping constant $t_{1}$ will be set as the energy unit $t_{1}=1$. By setting proper parameters $t_{2}$, $b$, we can get one or two nodal lines.

\textit{Interaction Hamiltonian:} In order to investigate the effects of interactions in the  semimetals, we use the Hubbard model
\begin{eqnarray}
H=H_{0}+U\sum_{i}n_{i,\uparrow}n_{i,\downarrow}-\mu \sum_{i}n_{i,\sigma},
\label{model}
\end{eqnarray}
where $H_{0}$ denotes a tight-binding Hamiltonian of the non-interacting semimetals as we have introduced above. $U$ is the on-site Coulomb interaction, $n_{i,\sigma}$ is the particle-number operator on site $i$ with spin $\sigma$ and $\mu$ is the chemical potential which will be fixed $\mu=\frac{1}{2}U$ at half-filling as the band structures hold particle-hole symmetry.

\textit{Numerical Method:} A good way to analyze the Hubbard model numerically is the exact diagonalization (ED). However, due to the limitation of computer memory capacity and speed of CPU, the solvable size of the lattice cannot be large. Thus one will not get enough momentum points to study the $k$-space distribution of the spectral function. The CPT which is based on the ED is another numerical method to solve the Hubbard model\cite{Senechal,Senechal1,Senechal2}. The Green's function in a finite size of lattice is first calculated through the ED. Then we can divide the whole lattice into many small lattices of this size which are usually called clusters. The inter-cluster hopping is treated perturbatively and the Green's function of the physical system is obtained from strong-coupling perturbation theory. As we get enough momentum points, we can track the evolution of WPs or nodal lines controlled by the variation of $U$.

\begin{figure}[htb]
	\begin{center}
		\includegraphics[width=0.45\textwidth]{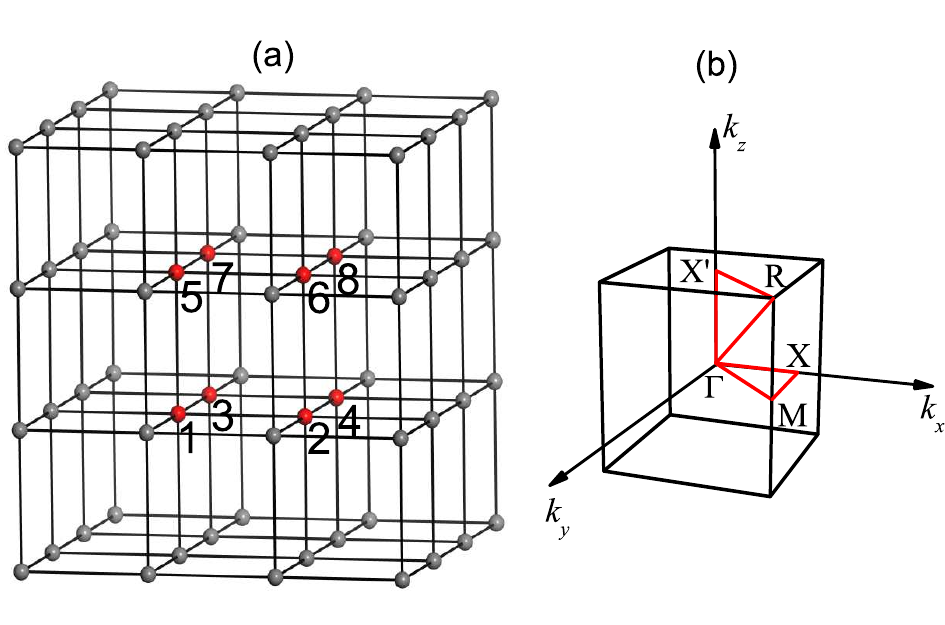}
		\caption{(a) A $2\times2\times2$-site cluster, as marked by numbers 1-8, in the cubic lattice. (b) The scanning routine is $\Gamma$-X-M-$\Gamma$-R-X$^{\prime}$-$\Gamma$ in the bulk BZ.}
		\label{fig1}
	\end{center}
\end{figure}

During our calculations, we choose a $2\times2\times2$-site cluster in the cubic lattice shown in Fig. \ref{fig1}(a). From the exact diagonalization, the Green's function in each cluster can be calculated from its definition
\begin{equation}
\begin{aligned}
G'_{\mu\sigma,\nu\sigma'}(\omega)&=\langle\Omega| c_{\mu\sigma}\frac{1}{\omega-H +E_{0}+i\eta}c^{\dag}_{\nu\sigma'}|\Omega\rangle \nonumber \\
&+\langle\Omega|c^{\dag}_{\nu\sigma'} \frac{1}{\omega+H -E_{0} +i\eta}c_{\mu\sigma}|\Omega\rangle,
\end{aligned}
\end{equation}
where $\mu$ and $\nu$ denote different lattice sites within a cluster with spin index $\sigma$,$\sigma'$ and $E_0$ the energy of the ground state. The system Green function $\mathbf{G}$ is obtained from the cluster Green function $\mathbf{G}'$ using the random phase like approximation
\begin{eqnarray}
\mathbf{G}(\tilde{\mathbf{k}},\omega)=\mathbf{G}'(\omega)
[1-\mathbf{V}(\tilde{\mathbf{k}})\mathbf{G}'(\omega)]^{-1},
\end{eqnarray}
where $\tilde{\mathbf{k}}$ belongs to the reduced Brillouin zone corresponding to the superlattice and $V_{\mu\sigma,\nu\sigma'}(\tilde{\mathbf{k}}) = \sum_{\mathbf{R}} V^{0\mathbf{R}}_{\mu\sigma,\nu\sigma'}e^{i\tilde{\mathbf{k}}\cdot\mathbf{R}}$ with $\mathbf{R}$ the superlattice index. $V^{0\mathbf{R}}_{\mu\sigma,\nu\sigma}$ is the hopping constant from sub-lattice site $\mu$ with spin $\sigma$ to site $\nu$ with spin $\sigma'$ between the original superlattice 0 and superlattice $\mathbf{R}$.

The $k$-dependent Green function $G^{\rm cpt}(\mathbf{k},\omega)$ is given by
\begin{equation}
G^{\rm {cpt}}_{\sigma\sigma'}(\mathbf{k},\omega)=\frac{1}{L}\sum_{\mu\nu}e^{-i\mathbf{k}\cdot(\mathbf{r}_{\mu}-\mathbf{r}_{\nu})}
G_{\mu\sigma,\nu\sigma'}(\tilde{\mathbf{k}},\omega),
\end{equation}
where $\mathbf{k}=\tilde{\mathbf{k}}+\mathbf{K}$ with $\mathbf{K}$ the reciprocal vector of the superlattice.

After getting the system Green's function by CPT, we study the spectral function of quasiparticles which can be obtained by
\begin{equation}
A(\mathbf{k},\omega)=-\sum_{\sigma}\mathrm{Im}G^{\rm cpt}_{\sigma\sigma}(\mathbf{k},\omega)/\pi.
\end{equation}
We have checked the results by varying the cluster size and found no qualitative difference.

\section*{Results and Discussion}

\begin{figure}
	\begin{center}
		\includegraphics[width=0.45\textwidth]{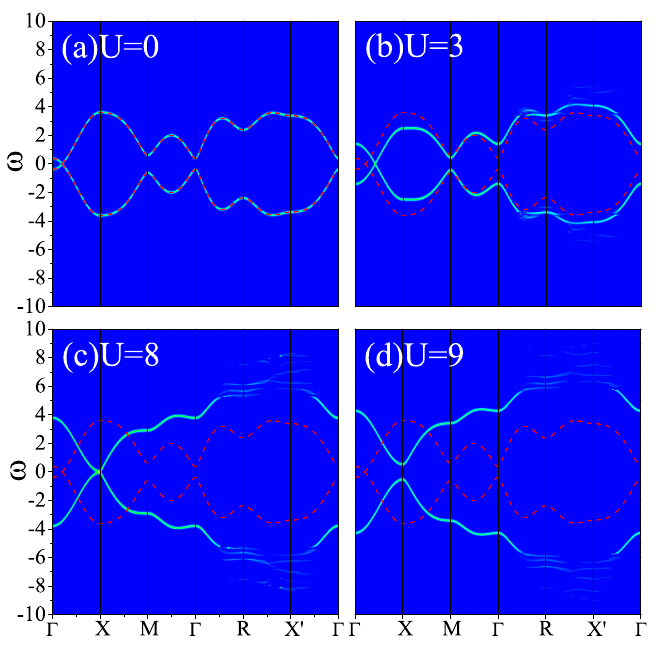}
		\caption{Intensity map of the spectral function $A(k,\omega)$ along high symmetry lines (see Fig.\ref{fig1}(b)) with $m=1.5$ and $k_{0}=0.2\pi$ based on the WSM1 model for (a) $U=0$, (b) $U=3$, (c) $U=8$ and (d) $U=9$, respectively. The red dashed lines denote the non-interacting band structures.}
		\label{fig2}
	\end{center}
\end{figure}

As for the WSM1 model, where the number of WPs is determined by the parameter $m$, we first set $m=1.5$ so that the system has only two WPs which exist at $(\pm k_{0},0,0)$. Intensity plots of $A(\textbf{\textrm{k}},\omega)$ along high symmetry lines are shown in Fig.\ref{fig2}, which gives a clearly evolution of the WPs for different interactions, as the WPs are initialized at $(\pm 0.2\pi,0,0)$. For the non-interacting state ($U=0$, Fig.\ref{fig2}(a)), the lower and upper band touch at the WP. When we increase interaction strength $U$, the WP shifts along the $k_{x}$ direction towards $X$ point and no gap exists in this state. Finally the WPs meet at $X$ point when $U$ reaches 8 with no band gap. As $U$ continues growing, a gap will be opened at $X$ point, e.g., $U=9$ shown in Fig.\ref{fig2}(d). The magnitude of the gap will also rise with $U$ and the system becomes an insulator.

\begin{figure}[htb]
	\begin{center}
		\includegraphics[width=0.48\textwidth]{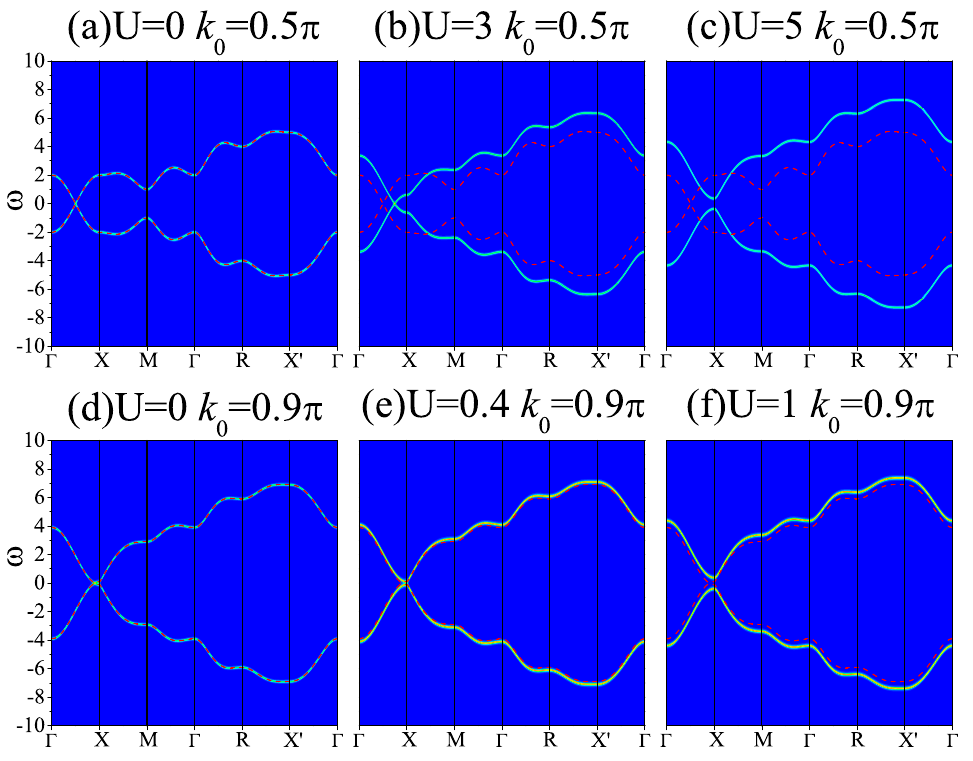}
		\caption{Intensity map of the spectral function $A(k,\omega)$ along high symmetry lines based on the WSM1 model: (a)-(c) $k_{0}=0.5\pi$ and $U=0$, 3, 5, respectively; (d)-(f) $k_{0}=0.9\pi$ and $U=0$, $0.4$, $1$, respectively. The red dashed lines represent the non-interacting band structures.}
		\label{fig3}
	\end{center}
\end{figure}

We then change the value of $k_{0}$ to see whether this phenomena is a special case. The results are shown in Fig. \ref{fig3} for $k_{0}=0.5\pi$ (a-c) and $k_{0}=0.9\pi$ (d-f). We can see that the WPs are pushed to the $X$ point and meet together by increasing $U$. After that a gap comes out and its magnitude increases with $U$. The difference among the results is that the nearer $X$ point the WP is , the smaller $U$ the gap opens for.

Another parameter we can change in this model is $m$, which affects the bandwidth and number of the WPs. If we increase $m$, the number of WPs will not change and the bandwidth will become larger. The movement of WPs is just the same under the present of $U$, as no qualitative change can be observed in contrast with $m=1.5$. If $m$ is decreased, the number of WPs will change to 6 or 8. The movement of WPs is similar under the present of $U$.

One possible explanation of the above movement of WPs is due to the mean-field analysis in the weak-interaction regime, as discussed in Refs. ~\cite{William,Acheche}. Specifically, the Hubbard interaction can be decoupled as
\begin{eqnarray}
Un_{i,\uparrow}n_{i,\downarrow}\rightarrow \frac{U}{2}(n_{i,\uparrow}+n_{i,\downarrow})-Um_xc^{\dag}_i\sigma_{x}c_i+Um_x^2,
\label{mf}
\end{eqnarray}
where $m_x=\frac{1}{2}\langle c^{\dag}_i\sigma_{x}c_i\rangle$ represents the magnetization order parameter along the spin-$x$ direction. For Eq.\ref{model1}, there is already a ferromagnetic moment $2(m-t\cos k_{0})\sum_ic^{\dag}_i\sigma_{x}c_i$ at $U=0$, where $2(m-t\cos k_{0})>0$ within our calculations for the two WPs case. At the mean-field level, the magnetization is enhanced by turning on the repulsive $U$, i.e., $-Um_x>0$ increases with $U$. Therefore, we see that the two WPs move to larger $k_0$ (e.g., $\cos k_0\rightarrow\cos k_{0}+Um_x$) with increasing $U$.

\begin{figure}
	\begin{center}
		\includegraphics[width=0.48\textwidth]{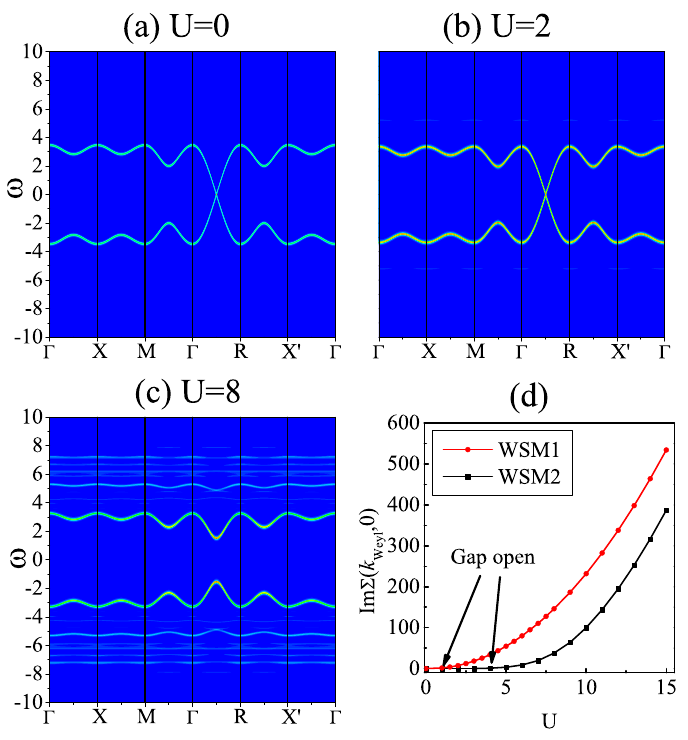}
		\caption{Intensity map of the spectral function $A(k,\omega)$ along high symmetry lines based on the WSM2 model for: (a) $U=0$; (b) $U=2$; (c) $U=8$, respectively. (d) The imaginary part of the self-energy at the Weyl points for WSM1 ($k_{0}=0.9\pi$) and WSM2 models when $\omega=0$. }
		\label{fig4}
	\end{center}
\end{figure}

We now turn to the WSM2 model. In this model, there are eight WPs. When the interaction $U$ is introduced, the positions of WPs do not change and a gap is opened at the WP when $U$ is large enough, as shown in Fig. \ref{fig4}. A possible explanation is again due to the mean-field argument: Since there is no magnetization at $U=0$, turning on a weak repulsive $U$ would not induce a magnetization at the mean-field level, e.g., $Um_x=0$, and hence the positions of the WPs are not affected. On the other hand, the stability of WPs under the Hubbard interaction (below a critical $U$) may be protected by the space-inversion symmetry of this model.

Now, a few remarks are in order concerning the nature of the metal-insulator transition in our WSM models \cite{Acheche}. The gap in WSM1 model opens after the WPs merge at the $X$ point, especially when the non-interacting WPs are very close to each other, a small $U$ could open the gap. While in the WSM2 model, the gap directly opens at the WPs until $U$ is large enough (see Fig. \ref{fig4}). For large $U$ the insulating phase should be a Mott insulator, which can be identified by checking whether the imaginary part of the self-energy at low frequency diverges \cite{Acheche}. The numerical results of the imaginary part are shown in Fig. \ref{fig4}(d). We see that, for the WSM2 model, the imaginary part grows very fast after the opening of the gap at $U=4$ (we do not see a sudden divergence due to the finite-size effect). This divergence indicates that the gap should be a Mott gap for the WSM2 model. Whereas for the WSM1 model, the imaginary part of the self-energy has no divergence when the gap opens, and this behavior persists within a large range of $U$. Obviously, this is not a Mott gap. When $U$ reaches about 4, it starts to diverge rapidly, which means that the system enters the Mott insulating phase. 
 
\begin{figure}
	\begin{center}
		\includegraphics[width=0.48\textwidth]{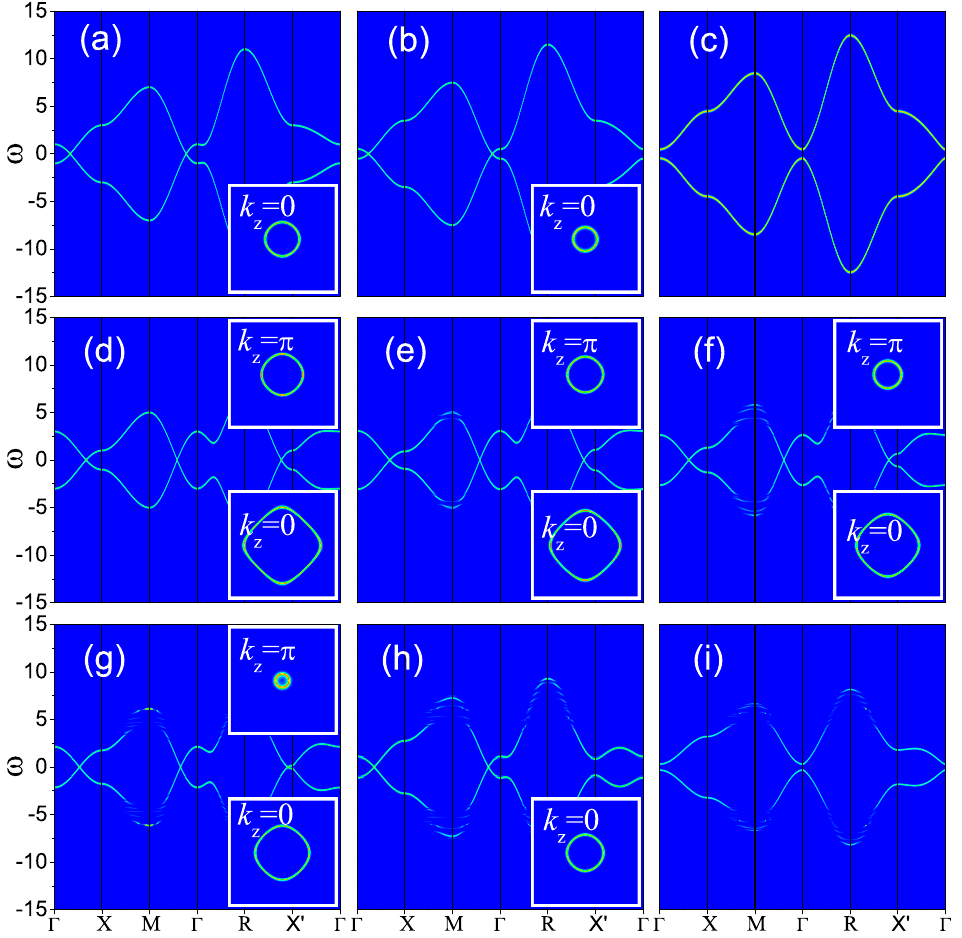}
		\caption{Intensity map of the spectral function $A(k,\omega)$ along high symmetry lines based on the NLSM model. Evolution of $A(k,\omega)$ in the one-nodal-line state is shown in (a)-(c) with parameters $t_{2}=1, b=1.5$ and (a) $U=0$; (b) $U=1$; (c) $U=3$, respectively. (d)-(i) show the evolution of $A(k,\omega)$ in the two-nodal-line state with parameters $t_{2}=0.5, b=0.3$ for (d) $U=0$; (e) $U=1$; (f) $U=2$; (g) $U=3$; (h) $U=5$; (i) $U=8$, respectively. The insets show the Fermi surfaces in $k_{z}=\pi$ or $k_{z}=0$ planes. }
		\label{fig5}
	\end{center}
\end{figure}

In the following, we will discuss the NLSM model. This model contains three adjustable parameters $t_{2}$, $t_{3}$ and $b$. One of them, $t_{3}$, will not affect the positions of nodal line and will be set $t_{3}=1$ during our calculations. $b$ controls the shape of nodal line and $t_{2}$ determines the number of nodal lines. After choosing suitable parameters, we get one-nodal-line and two-nodal-line states. We first set the parameters $t_{2}=1$ and $b=1.5$ under which the system enters the one-nodal-line state. We track the evolution of $A(k,\omega)$ by changing the interaction strength $U$. For the non-interaction state, we can see a nodal line in the $k_{z}=0$ plane and the upper and lower bands touch each other along the line (shown in Fig. \ref{fig5}(a)). As $U$ increases, e.g., $U=1$, we can see clearly in Fig. \ref{fig5}(b) that the position of the nodal line changes and the nodal line starts to shrink to the $\Gamma$ point. There is no band gap and the system remains as a semimetal. With a further increase of $U$, the nodal line continues its shrinking and finally become a point at (0,0,0). After that, if we keep increasing $U$, a gap will be opened at the $\Gamma$ point and the system become an insulator (see Fig. \ref{fig5}(c)). The evolution of $A(k,\omega)$ shows very similar as what we have discussed in the WSM1 model with two WPs.

Next we will discuss the results for the existence of two nodal lines. The parameters are set $t_{2}=0.5$ and $b=0.3$. There are two nodal lines: one lies in the $k_{z}=0$ plane, while the other lies in the $k_{z}=\pi$ plane. For the non-interaction state, shown in Fig. \ref{fig5}(d), we can see the nodal lines clearly. When an interaction $U$ is introduced in this system, such as $U=1$, 2 and 3, one nodal line lying in the $k_{z}=\pi$ plane begins to shrink toward $X'$ point, while the other lying in the $k_{z}=\pi$ plane shrinks toward $\Gamma$ point (see Fig. \ref{fig5}(e), \ref{fig5}(f), and \ref{fig5}(g)). As $U$ continues growing, the nodal line in the $k_{z}=\pi$ plane becomes a point at $X'$ point, and then a gap opens at the this point. However, the system remains as a semimetal, as the nodal line in the $k_{z}=0$ plane still exists. When $U$ is further increased, the nodal line in the $k_{z}=0$ plane continues its shrinking and the magnitude of the gap at $X'$ point increases, as shown in Fig. \ref{fig5}(h) with $U=5$ for example. When $U$ reaches around 8, both of the nodal lines disappear and a full gap is opened. The system finally goes into the insulating state.

The positions of nodal lines in the momentum space can be determined by equations $\cos k_{x}+\cos k_{y}=b$ for $k_{z}=0$ plane and $\cos k_{x}+\cos k_{y}=b+2t_{2}$ for $k_{z}=\pi$ plane. When $U$ is introduced into this system, the parameters $t_{2}$ and $b$ will be renormalized to larger magnitudes. As a result, the nodal lines will shrink to $X'$ point in the $k_{z}=\pi$ plane and $\Gamma$ point in the $k_{z}=0$ plane. Moreover, due to the extra parameter $2t_{2}$, the nodal line in the $k_{z}=\pi$ plane shrinks faster than that in the $k_{z}=0$ plane. To be more precise, let us elaborate this using the mean-field analysis, e.g., Eq.\ref{mf}. As can be seen from Eq.\ref{model3}, there is already a ferromagnetic moment $(-t_1b-t_2)\sum_ic^{\dag}_i\sigma_{x}c_i$ at $U=0$, where $(-t_1b-t_2)<0$ within our calculations for the NLSM model. At the mean-field level, the magnitude of magnetization is enhanced by turning on the repulsive $U$ (i.e., $Um_x>0$ increases with $U$), which effectively enlarge the parameters $t_{2}$ and $b$ as: $b\rightarrow b+Um_x, b+2t_2\rightarrow b+2t_2+Um_x$.

Finally, we would like to discuss the effect of quasi-particle weight $Z$ which is ignored in our mean-field analysis \cite{Acheche}. In the presence of the Hubbard interaction, the single-particle band structure is renormalized by the factor $Z$. Within the framework of mean-field theory, the low-energy effective Hamiltonian can be approximated as $H_{\textrm{eff}}=ZH_{\textrm{MF}}$, where $H_{\textrm{MF}}$ denotes the mean-field Hamiltonian obtained from Eq.\ref{mf}. Thus, we see that the quasi-particle weight $Z$ does not affect the positions of Weyl nodes and the form of nodal lines in our models.

\section*{Summary}
In summary, we have studied the evolutions of the WPs and nodal lines under interaction $U$ using CPT. For WSM1 model, the WPs move towards a specific point with the increase of $U$. When the WPs meet at the point, a gap is opened there and the system becomes an insulator. In WSM2 model, the WPs are static and when $U$ is strong enough, a full gap is opened at the WPs. In the NLSM model, we have discussed the one-nodal-line and two-nodal-line states. For the one-nodal-line state, the only nodal line shrinks to a specific point in its plane and finally a gap is open at the point. For the two-nodal-line state, the two nodal lines both shrink but not in step. When both nodal lines disappear, a full gap is opened with increasing $U$. We argue that the nontrivial evolution of WPs and nodal lines under interaction is due to the presence of symmetry breaking order, e.g., a ferromagnetic moment, in the framework of mean field theory, whereas the stability of WPs under interaction is protected by symmetry. Among all these models, the spectral gap is finally opened when the interaction is strong enough.

\section*{Acknowledgments}
The authors are very grateful to Jian-Xin Li and Yu Xin Zhao for useful discussions. This work was supported by the Fundamental Research Funds for the Central Universities of China (Grant No. 2013B00314) and the National Natural Science Foundation of China (Grant Nos. 11347111, 61404044, 11674158, 11704180 and 11704341).

\section*{Author contributions statement}

J.K. supervised the whole work, performed the numerical calculations and analyzed the data. J.Z., K.L., S.L.Y. and L.B.S. joined in the data analysis. All of the authors contributed to the writing of the manuscript.

\section*{Competing Interests}

The authors declare no competing interests.

\section*{Additional information}

\end{document}